# Towards Effective Cybercrime Intervention

Jonathan W. Z. Lim, Vrizlynn L. L. Thing
*Cyber Security Strategic Technology Centre, ST Engineering*
lim.weizheng.jonathan@stengg.com, vriz@ieee.org

*Abstract*—Cybercrimes are on the rise, in part due to technological advancements, as well as increased avenues of exploitation. Sophisticated threat actors are leveraging on such advancements to execute their malicious intentions. The increase in cybercrimes is prevalent, and it seems unlikely that they can be easily eradicated. A more serious concern is that the community may come to accept the notion that this will become the trend. As such, the key question revolves around how we can reduce cybercrime in this evolving landscape. In our paper, we propose to build a systematic framework through the lens of a cyber threat actor. We explore the motivation factors behind the crimes and the crime stages of the threat actors. We then formulate intervention plans so as to discourage the act of committing malicious cyber activities and also aim to integrate ex-cyber offenders back into society.

## I. Introduction

CYBERCRIME is no stranger to the law enforcement personnel, cyber defenders, or the public community. The recent COVID-19 pandemic has pushed for the transformation in the modes of communication, accelerated the culture of remote working and increased utilization of online collaboration platforms [1], [2]. These activities have resulted in the heavy reliance on digital and network communications, opening up new vulnerability channels. While the general population may still be navigating the changes and disruptions to their lifestyles, threat actors are constantly monitoring and thinking of exploitation avenues. Therefore, it is unsurprising to see a rise in cyber-attacks during this period [3]-[5].

The trend of rising number of cyber-attacks is not showing signs of slowing down, and it does not seem likely that we are able to eradicate cybercrimes either. In a 2020 Check Point Security Report [6], it was mentioned that the cyber landscape will continue to face challenges in Supply Chain, Cloud Environments, and Internet-of-Things devices. Ransomware and phishing attacks remain rampant, especially in the wake of the COVID-19 pandemic. Furthermore, the transnational nature of cyber-attacks makes it even harder to prevent and contain them. The impacts are borderless and widespread, often with devastating effects to the victims.

In dealing with cyber protection, many people understand the asymmetric nature between the role of a defender and that of an attacker. The defender has to consider and protect all vulnerable points of entry that can be exploited, while the attacker is able to focus on one or a few entry points to carry out an attack successfully [7]. As such, cyber defenders are inclined to develop technologies and solutions to provide the comprehensive protection to mitigate as many attacks as possible. This strategy of buffing the system defenses is indeed crucial to prevent breaches and damages, and may prevent the attacks from happening. However, it does not address the key question: Why are cybercrimes committed in the first place?

Our paper seeks to approach the crime prevention statement from another angle; to understand the life stages and features of a threat actor, the corresponding motivation factors that may be influencing their choices, and propose intervention plans that may be relevant to them. Kevin Mitnick, arguably the world's most infamous hacker, laments in his book *The Art of Deception* [8] that the human factor is security's weakest link. While the comment is mainly targeted at people who may be ignorant to security practices, it can also apply to threat actors, who may have extremist beliefs or have fallen into temptation to commit malicious activities despite knowing that they should not be doing so. With some proposed intervention plans, the hope is to discourage a life away from malicious activities, and integrate ex-cyber criminals back into society where they can contribute positively to the community.

The remainder of this paper is organized as follows. We explore a list of motivation factors that may be considered by threat actors in Section II. The crime periods, features, and interventions are discussed in Section III. We will conclude the paper with a reflection and outlook in Section IV, with thoughts on further research directions.

## II. Motivation Factors

Motivation factors form an important aspect when we deal with cybercrimes. Threat actors often carry out their activities with certain intentions and personal goals. In the case of Kevin Mitnick, it was his quest for personal knowledge and intellectual challenge [9], [10]. For the hackers belonging to Lazarus Group, it is believed that they were primarily motivated by financial gain, due to their assaults on numerous financial institutions such as banks from various countries [11]. No matter what the nature of the crime is, we can try to attribute a motivation factor, or at times, multiple factors to a cybercrime.

Numerous researches have been conducted to understand and classify cybercrime motivation factors. In [12], the authors identified six types of motivation factors; addiction to computer networks, curiosity in the networks, thrill of illicit findings, attaining power over computer systems, peer recognition from other hackers, and finding security vulnerabilities. Smith [13] categorized the motivations into four broad categories, namely economic, political, ideological and behavioral/emotional. In [14], Ablon mooted that threat actors are spurred by different motivations. Cybercriminals are generally motivated by monetary rewards, while cyberterrorists may be more motivated by ideologies. Li [15] explored and provided an extensive list of motivations in an effort to distinguish them as much as possible. However, though some of them are distinctive among one another, there are some that are still quite similar. Nonetheless, the author tried to highlight the differences to his best effort. This motivations list forms part of the basis in our proposed paper.

Through the review of these researches, we classify the motivation factors in a generic manner, under the umbrella of root motivational categories. For example, economic and monetary rewards may be grouped under financial gains. Hacking for entertainment and thrill of illicit findings may be grouped under thrill factors. Undermining a government website may be grouped under terroristic factors, and cyber robin hoods may be grouped under idealistic factors. This further classification helps to provide an overview of the root motivation factors of concern that can be relevant in each crime stage. For the crime stages, it is also possible that there are multiple root motivations that are associated with each stage. Additionally, some root motivations may play a bigger part in influencing the crimes committed.

Building upon the motivations mentioned in [15], we further categorize them by linking each motivation to a root category, or multiple categories in some cases. The following seven categories are adopted and used as root motivations to classify the list; they are for Financial Factor, for Reputational Factor, for Idealistic Factor, for Terroristic Factor, for Retaliation Factor, for Thrill Factor, and for Intellectual Factor.

Financial Factor: Goldman and McCoy [16] described that financial motivated attacks account for a considerable proportion of cyber-attacks. The main incentive for these attacks is to monetize cyber activities such as exploiting stolen credit card details, or selling personal identifiable information. It can also mean stealing intellectual property such as technologies and trade secrets of organizations, or gaining unauthorized access to services without having to pay for them.

Reputational Factor: Some threat actors view illegal cyber activities as a pathway to increase their online reputation. Through these activities, they gain a sense of popularity and recognition from their peers as a validation of their capabilities [17]. Additionally, they may adore and strive to emulate threat actors with a higher degree of reputation.

Idealistic Factor: This category of motivations arises from situations that do not conform to the personal beliefs and expectations of the threat actors. Government websites may be defaced due to the unhappiness of the public [18], and school systems may be hacked by their own students to change the examination grades [19]. Even though the self-justification to themselves is for a greater good or self-righteousness, the acts are conducted through unlawful means.

Terroristic Factor: In 2016, Ardit Ferizi was charged with cyberterrorism by the US Department of Justice [20]. He was caught hacking and obtaining sensitive information on US military and federal personnel, and subsequently shared the information with the ISIS. In [21], some threat actors were found to send death and sexual threats to their victims, in a bid to intimidate them to abide to their instructions. Such motivations can lead to devastating consequences to individuals and/or to the whole nation.

Retaliation Factor: For motivations belonging to this category, the victims initiate the cybercrimes themselves against the harm they received from cyber attacks and/or other forms of mistreat. In [15], the author cited a few examples where the victims reacted with unlawful actions against their perpetuators. In [22], Lewis described that there were numerous voices in the US government to allow organizations to retaliate against cyber attacks, and added on to remark that it was not a good idea.

Thrill Factor: For some incidents, the purposes of them cannot be determined, even after going through the investigations. Some threat actors target their victims for no specific reason, and often just for the thrill of it [23]. This may occur when the threat actors wish to experience the satisfaction and adrenaline of hacking into the systems successfully, or simply because they find it enjoyable or the act to be humorous. Although the incidents may not directly cause any immediate financial loss as the threat actors are not motivated by such, it will nonetheless, adversely affect the image and branding of the victims. Eventually, this lowers the confidence from investors and customers alike, leading to reduced businesses.

Intellectual Factor: Hacking for intellectual reasons is common both in ethical and non-ethical hacking communities. Ethical hackers invade systems to discover vulnerabilities, or to test novel techniques and improve their penetration testing skills. All these are done in a controlled fashion, and usually initiated by the legitimate parties hiring them. For non-ethical hackers, targeting companies with shored up cyber defenses is one of the best ways to assess, gauge and prove their capabilities. The main motivation is their quest to improve their skills and intellectual capabilities, with little regards for moral values.

Table I highlights our categorization of the motivations, based on how each of them is associated to the root motivation(s). This presents a clear overview to where each motivation belongs. Additionally, in the list of motivations as described in [15], the author raised the possibility that some motivations can be uncertain, or affected by mental challenges. For such cases, it may be difficult to associate a root motivation with them, as the justification to commit the act may not be straightforward. For this reason, we have excluded them from our table. These excluded motivations are labeled as *Unclear motive* (e.g. perpetuator accessed and printed unauthorized information about an acquaintance, and subsequently handed a copy of the information to the victim himself) and *Influence by psychological depression* (e.g. perpetuator was diagnosed to have committed the act under the state of clinical depression and unsound mind).

Looking at the categorization, we can see that there is a wide range of root motivation factors that can affect the decision making of the threat actors to achieve their causes, or to satisfy their own desires. As such, we will survey the environmental and individualistic features that can potentially guide the threat actors to rightful or wrongful directions.

TABLE I
MOTIVATIONS LISTED AND CATEGORIZED ACCORDING TO ROOT MOTIVATIONAL FACTOR(S)

| Motivation | Financial Factor | Reputational Factor | Idealistic Factor | Terroristic Factor | Retaliation Factor | Thrill Factor | Intellectual Factor |
|---|---|---|---|---|---|---|---|
| Pursue free flow of information | | | ✓ | | | | |
| Realize free expression of ego | | ✓ | | | | | |
| Take technical challenge | | | | | | ✓ | ✓ |
| Curiosity to seek new knowledge | | | | | | | ✓ |
| Test system security and resilience | | | | | | | ✓ |
| Take an adventure in cyberspace | | | | | | ✓ | |
| Practice and show off programming and technical skills | | | | | | | ✓ |
| Tentative attacks against potentially vulnerable devices | | | | | | | ✓ |
| Hacking out of hatred | | | | ✓ | | | |
| Hacking to acquire financial gains or avoid payment | ✓ | | | | | | |
| Hacking into more advanced systems to leverage on their capabilities | | | | | | | ✓ |
| Hacking to change academic results | | | ✓ | ✓ | | | |
| Harassment and/or murder | | | ✓ | ✓ | | | |
| Mobilize political movement | | | ✓ | ✓ | | | |
| Launch cyber warfare | | | | ✓ | | | |
| Unleash anti-computerization actions | | | | ✓ | | | |
| Cyber robin hoods | | | ✓ | | | | |
| Create unfair competition | ✓ | | | | | | |
| Execute trap marketing | ✓ | | | | | | |
| Self-defense | | | | | ✓ | | |
| Hacking for recreation | | | | | | ✓ | ✓ |
| Employment-related motivations | ✓ | ✓ | | | | | |
| Defend the unethical hackers community against legal measures targeted upon them | | | | | ✓ | | |
| Destroy evidence present in information systems | | | | | ✓ | | |
| Sexually motivated misuse | | | | ✓ | | | |
| Deviance | | | | | ✓ | | |

### III. CYBERCRIME TIMELINE STAGES

Many proposed models seek to associate motivation factors with cybercrimes to understand the purposes behind. Law enforcement personnel or cyber defenders can leverage on these models to learn about the threat actors, their motives and operating habits to better identify them. Other than building the criminals' profiles, the models can also aid the development of cyber defenses based on how the threat actors may attack their targets. Organizations are able to develop defense strategies based on the potential motivational factors, and prevent the attacks from being executed successfully. It may not be easy or possible to develop a one-size-fits-all approach, but the impacts can still be mitigated. In the following sub-sections, we will look at the existing models and tools that are available, and on our proposed methodology to address challenges of cybercrime.

*A. Existing Models and Tools to Address Challenges of Cybercrime*

Modarres [24] proposes a model based on three root factors that can influence a cyber-attack. By integrating the three root nodes, the authors form an influence model to determine the nature of the attack. They further break down the nodes to finer granularity, establishing sub-nodes that may affect the reasoning to the root motivations. The model also allows different types of links between nodes and sub-nodes to highlight various influences such as amplifying effect, negative connotation, and the direction of influence that the nodes have on one another. The model is shown in Fig. 1.

- Motivation: Describes the inclination of a threat actor based on the perceived benefits.
- Opportunity: Describes the circumstances and conditions that are advantageous to a threat actor.
- Deterrence: Describes the elements that will discourage a threat actor from committing a malicious act.

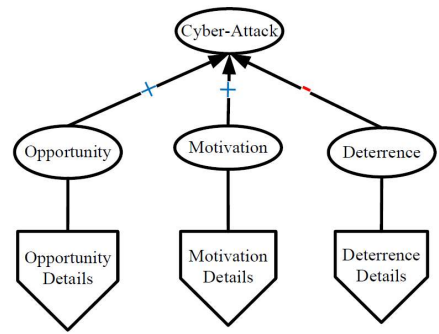

Fig. 1 The Influence Model encompassing root nodes for a cyber-attack [24]

In Ngafeeson's proposed model [25], two key motivational theories are discussed to create a motivational framework for cybercrime classification. These two theories are Maslow's Theory of Hierarchical Needs and Herzberg's Motivation-Hygiene Theory, which focus on the needs and supporting elements to sustain a human physically and psychologically. They form the motivation component as part of the overall framework to explain how potential threat actors can be driven by different needs, employ the use of technologies, and bypass security barriers to reach their security targets. The model aims to provide a holistic and comprehensive perspective to crime classification, combining the elements of humans such as victims/perpetuators, and the elements of tools and systems, such as technologies involved and security barriers. The model framework is depicted in Fig. 2.

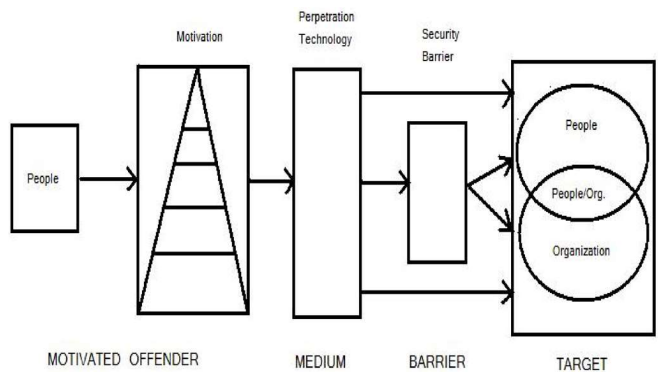

Fig. 2 The Motivational Model of Cybercrime Classification in [25]

Dupont [26] advocated for a monitoring platform to enhance the support and response to crimes. He believed that it is important to harmonize the efforts in tackling cybersecurity crimes, and measure the outcomes and effectiveness of the policies and programs through a consolidated platform. The platform will facilitate the sharing of information, ideas and policies that can provide solutions and knowledge to users who may be facing similar conditions. In the proposed platform, the author designed a pilot version of the coding framework with multiple search fields for data as shown in Fig. 3. The aim is to collect as much information as possible, according to the data fields designed in it. The data fields are well thought through, consisting of searches related to different categories such as policy information, evaluation types, methodologies, and any supporting documents that may be useful. Other than just facilitating the knowledge database, the platform may also bring forth other advantages such as tracking the effectiveness and ineffectiveness of the policies. Both features are equally important to allow the review of policies, and enable the refreshing of data if they are outdated or fail to produce the intended results.

The mentioned models and tools are useful for discussing plausible motivation factors that may drive cybercrimes. However, there is a lack of correlation between the motivation factors and any conceivable plans to address the motives discussed.

| Categories | Data Fields |
|---|---|
| Overview of the policy and search filters | Summary |
| | Nature of the policy |
| | Related policies and legislation |
| | Keywords |
| | Snapshot data |
| Description of the policy | Date of implementation or launch |
| | Place of implementation |
| | Geographical scope |
| | Instigator of the policy |
| | Targeted issue or situation |
| | Targeted population |
| | Goals of the policy |
| | Components of the policy |
| | Agents in charge of implementation |
| | Costs |
| | Source of funding |
| | Penalties |
| | Incentives |
| | Challenges |
| | Implementation information |
| Evaluation of the policy | Existence of an evaluation |
| | Evaluation type |
| | Evaluator |
| | Methodology |
| | Outcomes |
| Additional references | URL |
| | Peer-reviewed publications |
| | Media articles |
| | Official documents |

Fig. 3 Coding Framework of Monitoring Platform in [26]

Our paper aims to address this lack of correlation from a fresh angle, and approach this through the lens and life stages of a threat actor. This angle is discussed in the next sub-section.

*B. Proposed Methodology through the Cybercrime Stages of a Threat Actor*

The proposed methodology builds upon the timeline pertaining to the cybercrime stages of the threat actor. They are divided into three segments, namely *Pre-crime period*, *Crime period* and *Post-crime period*. Using this timeline and the root motivational factors as foundations, we postulate the possible environmental and individualistic features, which may affect potential threat actors' decision whether to initiate crimes and go through with the entire process. These features may be unique to one period, or may be applicable to multiple periods. If they are applicable to multiple periods, the presumption is to classify them at the period where they may have the strongest impacts. This will help to provide some insights on the potential challenges faced by them at each stage. Through this understanding of the conditions and motivation factors, intervention plans are proposed to tackle these challenges.

For each stage in the timeline, we will explore different aspects that can influence the actor's decision to commit a crime. First, we look at the environmental features as external influences, such as the behavior of peers around the threat actor, the characteristics of cyber space or the social mindset pertaining to ex-convicts. We think of these conditions as external influences being inflicted upon the actor. Next, we consider the individual features such as the emotional response towards punishment, the self-perceived level of the actor's capabilities, the state of mind of the actor, or any other psychological activators that may trigger a reaction. We think of these conditions as internal influences within the actor. Last but not least, we consider the availability of knowledge and resources that the actors can obtain from, so as to learn the necessary skills and enable them to perform the crimes.

Pre-crime period: This period refers to the time span where the person has not committed any crime, but may evolve to become a potential threat actor. During this period, the activities that the actors engaged are seemingly legitimate. However, they may be facing a great deal of challenges and temptations. For example, avid gamers are often well aware of game modifications or cheat codes, and can use them to their advantage if they are willing to resort to such unorthodox means. Such acts will breach the terms of the games, and gamers risk account banning. However, prosecution by law is still debatable for such incidents [27]. Coupled with generic leniency for first-time offenders, it does not help much in dissuading them from repeated offences, much less eradicate their wrongful intentions. Even so, this is also dependent on the nature of the crime. For example, in [28], the author discussed the arguments either in favor of, or against a reduced sentence for first time offenders. If the first-time offenders are of young age, they may lack maturity and discernment of the resulting consequences due to their actions. Thus, there is a stronger plea to reduce the punishment. However, if the first offence is a rape incident as compared to a shoplifting incident, it is arguably trickier to advocate for a reduced sentence.

Another potential challenge is that the person is currently on the receiving end of cyber pranks or attacks. If the effect is damaging to a level that the person cannot withstand, they may seek revenge against the perpetuators, even if it means overstepping the legal boundaries [29]. There are also other temptations such as the opportunity of making a small but quick buck illegally, with perceived low risks [30]. Often, it starts with a small amount, but leads to numerous attempts in the future for greater financial reward, due to greed and further perception of low risk due to accumulated successes.

The anonymity of the cyber realm and activities carried out within, enhances the person's motivations and intentions. Thinking their wrongdoings will not be detected and they will not be caught, it furthers their cause, and provide them the impetus to execute their first crime.

Based on the features in the pre-crime period, the common root motivations gravitate towards the factors of idealistic, terroristic retaliation, thrill and intellectual. They may have dabbled into illegal activities due to their curiosity to experiment new security techniques, or accidentally ventured into them without knowing. Although possible, financial and reputational factors at pre-crime period are less likely to be the main and sole motivations as cyber attacks are often no easy feat, and potential threat actors have other legitimate opportunities and means to secure their income and build their portfolios. The features of the pre-crime period are highlighted in Table II.

| *Pre-crime Period* ||
|---|---|
| **Environmental Features** ||
| 1 | Perceived non-illegal hacking outlets (e.g., game modifications, cheat codes) |
| 2 | Anonymity of cyber realm and activities |
| 3 | Perceived reduced punishment for first offence |
| **Individualistic Features** ||
| 1 | Clean slate record |
| 2 | Victims of cyber attacks |
| 3 | Temptation to make a quick buck |
| **Knowledge Resources** ||
| 1 | Formal education (e.g., computer science, computer engineering) |
| 2 | Self-help online resources |

TABLE II: Features of the pre-crime period

Crime period: This period refers to the time span where the threat actor has committed a crime, but not arrested by law enforcement yet. During this period, the threat actors face a new set of challenges through their first misconduct. The key focus here is to review the features that may push them to commit their second crime and onwards, or prompt them to repent and acknowledge their mistakes. In the case of the former, threat actors may perceive these acts as of a non-violent nature, justifying to themselves that they did no physical harm to the other human beings. Additionally, threat actors may have obtained a heightened sense of confidence, gaining trust in their own capabilities and methods to commit crimes and go undetected, after the successful first attempt. Finally, if their motivation was to seek a thrilling experience, they may indulge in competing with fellow hackers to boast about their misguided achievements, without much consideration to the consequences of their actions. However, if it is the latter, there are some features in this period that may prompt a response towards guilt and repentance. The thought of living an evasive life from the authorities and hidden life from friends and family may be too daunting for some to bear, as it also comes with the lingering fear of arrestment.

Based on the features in the crime period, the common root motivation factors gravitate towards the factors of reputational, idealistic, terroristic, retaliation, thrill and intellectual. As threat actors had a taste of illicit activities without condemnation, they may become more audacious in their behaviors to further their own causes. Similar to pre-crime period, financial factor may be of lesser concern as threat actors are still likely to have legal channels of income if their illegal activities are not yet exposed. These features are highlighted in Table III.

| Crime Period | |
|---|---|
| **Environmental Features** | |
| 1 | Unintentional illegal intrusion into unauthorized systems |
| 2 | Peer competitive spirit to boost reputation without consideration to consequences of hacking |
| 3 | Evasive life due to the need to manage both lawful and unlawful activities (e.g., private vs public working/living/network spaces of activities) |
| **Individualistic Features** | |
| 1 | Heightened sense of confidence after successful intrusion into systems without being detected |
| 2 | Lingering fear of sanctions or arrest |
| 3 | Perceived acts possibly as non-violent |
| **Knowledge Resources** | |
| 1 | Formal education (e.g., computer science, computer engineering) |
| 2 | Self-help online resources |
| 3 | Hackers community |

TABLE III: Features of the crime period

**Post-crime period**: This period refers to the time span where the threat actors have committed their crimes, and are caught by the authorities. It is applicable where they are caught after committing a single crime or multiple crimes. Here, we make the assumption that their sentences and punishment do not result in their deaths or life imprisonment, which means they have a second opportunity to regain their livelihood back in society in the future.

During this period, threat actors often face social stigmas and personal difficulties beyond their control. Due to their criminal records, they may face reduced job opportunities, as hirers may be concerned over offences that are potentially applicable to the organizations' setup and environment. The time spent in confinement, especially if the duration is very long, can also contribute to them losing touch with the society and skill relevancy to the industry. There may also be disruptions to their previous social network and support, making it even harder for them to integrate back to the society.

On the personal aspect, their technical and social skills may also deteriorate, partly because of the lack of practice and work experience. Since the incarceration, they may have also lost their financial support and assets. Faced with these difficulties, the likelihood of recidivism increases.

Based on the features in post-crime period, the common root motivation factors gravitate towards the factors of financial, terroristic and retaliation. Financial stability is the main concern due to the difficulty in landing jobs after they are released. Terroristic and retaliation factors can continue to be strong motivations as they are impelled strongly by personal beliefs. Table IV highlights the features in this period.

These three periods mark the evolution of potential threat actors to becoming real and/or serial criminals. For each crime stage, the conditions and motivations that they face may be distinct, or may be applicable to other stages as well. As such, based on these features and conditions, we propose relevant intervention plans to reduce cybercrimes and integrate ex-convicts back into society, which directly address the motive of our paper, in the next sub-section.

| Post-crime Period | |
|---|---|
| **Environmental Features** | |
| 1 | Reduced job opportunities possibly due to criminal record |
| 2 | Lost touch with society and skill relevancy to industry |
| 3 | Disrupted social network and support |
| **Individualistic Features** | |
| 1 | Deteriorated technical and social skills |
| 2 | Loss of financial support and assets after incarceration |
| 3 | Recidivism |
| **Knowledge Resources** | |
| 1 | Formal education (e.g., computer science, computer engineering) |
| 2 | Self-help online resources |
| 3 | Hackers community |

TABLE IV: Features of the post-crime period

*C. Proposed interventions*

There are many different strategies to address the topic of cybercrime prevention. One approach is to focus on building and fortifying the organization's defenses through the security products and solutions [6], [31]. This includes having security monitoring and detection systems in place, incorporation of security protection and real-time threat intelligence to prevent security breaches, manage and monitor access management controls, and other security tools and policies to eliminate security gaps. Another approach is to focus on the human behavior to shape their reactions when challenged with security obstacles [32]. Such areas can include raising security awareness to spot for malicious indicators, establishing good cyber hygiene practices and demonstrate appropriate reactions upon discovering cyber threats. As a holistic approach, organizations spare no effort to ensure their staff members are properly trained to be wary of the cyber pitfalls, and to shore up their cyber defenses against potential exploitations.

In terms of cybercrime prevention, these strategies can be viewed as a defensive stance in relation to the threat actor. This means that they do not actively tackle the motivations of the cyber attacks, which are no less important. Thus, in this paper, we are adopting the posture of seeking to encourage the change in mindset and behaviors of the potential threat actors, threat actors, and the public community pertaining to cybercrime prevention. These are accomplished via the proposed intervention plans, spanning distinctly in each crime period or across a combination of them. To the best of our knowledge, these intervention plans are not readily deployed globally. It is our hope that these proposed plans and framework can trigger more conversations to improve the measures on reduction of cybercrimes in a holistic manner. The interventions are intended to function as initial starting points and encourage afterthoughts, rather than to serve as hard-fix solutions.

The following plans are designed through the insights gained from the features highlighted in the different crime periods:

**[i] Formalize cyber topics into early childhood education and curriculum:** Education plays an utmost important role, either to learn skills for cyber development, to stay vigilant against threats, or to frame the mindsets of individuals. Technological advancement has allowed children to obtain digital devices at a progressively younger age. As the digital environment becomes more prevalent, the corresponding risks are also increasing. At a young age, humans are taught to be cautious of things that may pose a physical danger (e.g., fire, sharp objects). For items that are seemingly used for leisure purpose (e.g., phones, tablets), the dangers are not as obvious, and definitely not as equivalent to the physical danger.

The benefit of introducing cyber topics early is two-fold. First, it helps to build the knowledge and understanding of the consequences of cyber-attacks, the harm that they can bring to systems, and adverse impacts to humans' livelihoods. We assume that the majority of population is inherently good and do not wish to cause harm to others. Learning about the consequences caused by malicious actions may deter them from committing the act in the first place. Second, it is an opportunity to set healthy cyber pathways in motion. Such initiatives may entail learning on ethical channels of cyber activities and prospects in cyber careers, to inspire generations of cyber defenders.

The main aim of this plan is to institute a strong knowledge base of cyber topics, and inculcate a wholesome set of moral values. By molding the character from young, it builds the person and contributes to reshaping the cyber landscape (e.g., lowered cybercrime rates).

**[ii] Cultivate community service programs for first time minor offenders, as an alternative to prison term:** There are instances where first time offenders ventured misguided or unknowingly into a crime, often with no malicious intent. It may be due to moments of poor judgement, or inner mental traits that drove them there. They may not see far beyond their actions and the damages that they may cause. As such, the plan can be tailored to educate them on the repercussions of their actions and encourage them to be more aware of and admit to their mistakes.

Though it is with the belief that in a fair justice system, criminals should not be totally let off without consequences, and thus, sending the wrong message to them, their victims and the public, the degree of punishment can and should be administered with appropriate care. This is to also, fairly justify the purpose of the punishment to serve as a sufficient deterrent from future offences and at the same time, answer to everyone at stake for the committed offence. Instead of jail sentences, combinative alternatives may include volunteering activities to educate the public on cyber awareness, participate in cyber trial initiatives to examine their effectiveness, and undertake learning modules of cyber risks and impacts. In the long term, the absence of a criminal record may prove to be more impactful socially and mentally for the person, rather than its presence.

**[iii] Reconstruct the working boundaries for ex-offenders:** The thought of hiring ex-offenders is an ongoing debate with divided factions; there is no full swing towards supporting any side. The aspiration may be to welcome ex-offenders back into the working society, but the harsh reality is that many of them still face unemployment and discrimination. This is most evident in background screening during the hiring phase. Comparing two people with similar skillset but one of them having a criminal record, there is no impetus to select the ex-offender judging from a trust standpoint [33]. Additionally, the fear of recidivism is a legitimate concern for any employers. In the field of cyber security, this concern is amplified given that they may have the capability to infiltrate your establishment and overturn your security from the inside if hired. Nonetheless, they should not be deprived of an opportunity to redeem themselves.

One possibility is to explore and legislate the working configuration for ex-offenders where their past records will not hinder the hiring considerations as much. An example would be involving them in working with other internal teams to devise strategies to prevent and hunt for insider threats. Having the experience on the offensive side of security, they may provide insights on how to manage such threats and the thinking behind them. Another possibility is to recruit them to participate in technology research and development. Through this method, they can still be engaged intellectually to research and develop viable proof-of-concepts. As with network segregation for cyber protection, there are many ways to segregate roles and responsibilities, and identify important areas of work with good career growth opportunities, that are suitable for cyber ex-offenders.

**[iv] Raising the lower limit of maximum punishment for intentional and repeated offenders:** The landscape for the prosecution of cybercrimes is relatively newer and with relatively lesser precedents, in part due to the anonymity and transnational nature of them. Furthermore, details of cybercrimes can vary widely with not many similarities between each case. As a result, prosecutors may find it difficult to determine the appropriate punishment with no suitable reference for comparison [34].

Looking at the upper limit of the maximum punishment for cybercrimes, the UK's Computer Misuse Act prescribed the life imprisonment for offences relating to the loss of life or endangering national security [35], [36]. Under the US's Computer Fraud and Abuse Act, the maximum sentence is also life imprisonment for any offences that result in deaths [37]. In some countries, the maximum punishment can lead to death sentences depending on the nature and effects of the acts [38], [39]. The upper limit is seemingly more than adequate to offer deterrence, but the same cannot be said to the corresponding lower limit. Many of the sentences prescribe one year jail term as the lowest, which casts doubt on whether the measure is fully effective as a deterrence. In order to establish a harder initial stance to prevent the act in the first place, this aspect can be reviewed further.

**[v] Set up cyber rehabilitation centers to provide job transition opportunities, counselling assistance and spearhead community service programs:** In 2017, UK's National Crime Agency conducted its first rehabilitation camp

for young cyber criminals [40]. The main goal of the camp is to channel their competency towards healthy and lawful outcomes. Another goal is to assess their abilities for potential recruitment in the future. The idea of rehabilitation centers is often used for drug and alcohol addiction, but it can also be applicable to reforming cyber offenders. The rehabilitation camp mentioned is an example of a small-scale initiative. At a larger extent, cyber rehabilitation centers can also explore job opportunities for ex-offenders and provide support in encouraging their transition back into society. Within the centers, job opportunities could be present too, such as job scopes which encompass spearheading reformative initiatives for those in need, managing partnerships with key stakeholders, or to build in-house security capabilities.

*D. Mapping and integration of interventions with crime stages*

The next element of the framework is to relate the interventions with the three crime stages in a meaningful manner. Each intervention is mapped into either a crime stage, a combination of two crime stages, or three crime stages, if applicable. This is presented in Fig. 4. By associating them in this manner, we can have a more comprehensive outlook and clearer view of the relevance of intervention plans at each crime stage.

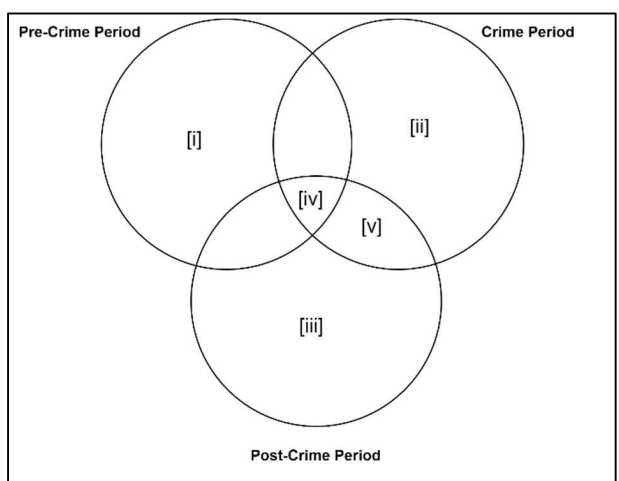

Fig. 4: Mapping the interventions onto crime stages

For example, intervention (i) may prove to be most meaningful when applied during the pre-crime period. The education received can influence the pathway during the formative years of the person. The main purpose is the aversion of a person from carrying out a cybercrime.

For intervention (ii), the objective is not to condemn the person but the crime, and to spur the apologetic behavior towards repentance. This is most relevant to unintentional first-time offenders, who do not wish to continue their missteps.

Intervention (iii) explores the avenues that ex-cyber offenders and businesses alike can take, when looking for talents in an increasingly cyber security demanding landscape.

In intervention (iv), the hope is to discourage any person from thinking of pursuing the crime life, any offenders from continuing to violate the law while risking getting caught, or any ex-offenders from intentionally committing crimes again. Thus, this intervention plan can be applied to all crime stages.

For intervention (v), offenders who had brushes with illegal activities can look for channels to confess while seeking guidance to reform themselves, or for ex-offenders who require assistance to integrate back into society which can be facilitated through various initiatives coordinated by the Center.

IV. REFLECTION AND OUTLOOK

In cybercrime prevention, many factors can affect the beliefs and decisions of the threat actors. Building up security defenses of your solutions and products remain an integral component of cyber security, as it can potentially deter ill intentions and prevent breaches and mishaps. This approach is leaning towards the adoption of technology to protect the systems, data, users and organizations. It is our belief that the other approach of tackling the social and mental construct of the threat actors, can work hand-in-hand in raising the effectiveness of reducing cybercrimes. For both approaches, resources from the government can greatly aid the launch of the proposed initiatives. When businesses see that the authorities are willing to walk the talk in lessening the number of incidents and helping ex-cyber offenders, they will also be more inclined to change their mindset and provide support as well. Thus, it takes commitment from all corners of the society.

For future research directions, we believe there will be more works on tracking the effectiveness of the various measures for cybercrime prevention. Our paper focuses on the framework and methodology of cybercrime prevention, and may serve as a reference to expand the studies in this topic. The measures and results require a long-term tracking, and more analysis can be done in the future to refine the intervention plans. We hope to see a heavier focus toward helping to shape the behaviors of threat actors or potential ones to reduce cybercrimes in the future.

**Jonathan W. Z. Lim** graduated with a Bachelor's degree in Electrical and Electronics Engineering from Nanyang Technological University, Singapore in 2013.

He is a Cyber Security Researcher at the Cyber Security Strategic Technology Centre, ST Engineering. He holds a general interest in cyber security, and has been involved in research relating to cyber governance and sustainability.

**Vrizlynn L. L. Thing** received her Ph.D. degree in Computing from Imperial College London, UK, while holding the UK Overseas Research Students Award and the Imperial College London, Department of Computing Scholarship concurrently. During her Ph.D. studies, she won the "Best Student Paper Award" at the 20th IFIP International Information Security Conference, and the Imperial College London "Hilfred Chau Postgraduate Award", presented to "students with exceptional achievements in scholarship", for her work on adaptive detection and mitigation of cyber attack mutations. She obtained her Bachelor's (Hons) and Master's degrees in Electrical and Electronics Engineering from Nanyang Technological University, Singapore, and her executive leadership training from National University of Singapore Business School.

She is the Senior Vice President, Head of Cybersecurity Strategic Technology Centre at ST Engineering. She provides thought leadership and guide her teams to lead and drive security technology innovation so as to enable the creation of differentiating products, solutions and services, and advise on strategic projects. She also holds the appointment of Honorary Assistant Superintendent of Police (Specialist V) at Singapore Police Force, Ministry of Home Affairs. Prior to joining ST Engineering, she was the Head of Cyber Security & Intelligence at A*STAR, Singapore. Her research on cyber security and digital forensics draws on her multidisciplinary knowledge in computer science and electrical, electronics, computer and communications engineering.